\documentclass{aastex}          
\usepackage{spr-astr-addons}    
\usepackage{natbib}             
\usepackage[hyperindex,breaklinks]{hyperref}
\usepackage{multirow}
\usepackage{bigstrut}
\usepackage{graphics}
\usepackage{amsmath}

\newcommand{\rmn}{\textrm}
\begin{document}

\title{Luminosity function of luminous compact star-forming galaxies}

\shortauthors{S. L. Parnovsky, I. Y. Izotova}

\shorttitle{
LF of luminous compact star-forming galaxies
}

\author{S. L. Parnovsky}
\affil{Astronomical Observatory of Taras Shevchenko Kyiv National University\\
Observatorna str., 3, 04053, Kyiv, Ukraine\\
tel: +380444860021, fax: +380444862191\\ e-mail:par@observ.univ.kiev.ua}
\email {par@observ.univ.kiev.ua}

\author{I. Y. Izotova}
\affil{Astronomical Observatory of Taras Shevchenko Kyiv National University\\
Observatorna str., 3, 04053, Kyiv, Ukraine\\
tel: +380444860021, fax: +380444862191\\ e-mail:izotova@observ.univ.kiev.ua}
\email {izotova@observ.univ.kiev.ua}

\begin{abstract}
We study H$\alpha$, far- and near-ultraviolet luminosity functions 
(LF) of the sample of 795 luminous compact star-forming galaxies with $z<0.65$. 
The parameters of optimal functions for LFs are obtained using the maximum 
likelihood method and the accuracy of fitting is estimated with the $\chi^2$ method.
We find that these LFs cannot be reproduced by the Schechter function because of an excess 
of very luminous galaxies. On the other hand, the Saunders function, 
the log-normal distribution and some new 
related functions are good approximations of LFs. 
The fact that LFs are not reproduced by the Schechter function can be 
explained by the propagating star formation.
This may result in an excess of luminous starbursts with the mass of  
a young stellar population above $2 \times 10^8$ M$_{\odot}$ as compared to
the LF of the quiescent galaxies. The most luminous compact galaxies 
are characterised by H$\alpha$ luminosities of 
$\ge 5\times 10^{42}$ erg s$^{-1}$ and star formation rates of 
$\ge 40$ M$_{\odot}$ yr$^{-1}$.
\end{abstract}

\keywords{Galaxies: luminosity function, mass function
--- Galaxies: starburst --- Galaxies: star formation}

\section{Introduction}\label{s:Introduction}

We study the luminosity functions (LFs) of the sample of 795
luminous compact galaxies (LCG) for the H$\alpha$ emission line, far ultraviolet (FUV) 
and near ultraviolet (NUV) continua. Recent bursts of star formation with age less than 
6 Myr are present in all galaxies as evidenced by very high H$\beta$ emission-line 
equivalent widths. Thus, our sample consists of extreme compact galaxies,
populating the high-end tail of the LF of the overall star-forming population. 

In this paper we consider five functions to
fit the LFs of our sample, namely the Schechter function \citep{ref:Schechter}, the Saunders 
distribution \citep{Sau}, the log-normal one and two
functions proposed by \citet{P}. The Schechter function was commonly used to approximate 
LFs in the H$\alpha$ emission line and ultraviolet (UV) range. 
However, it was argued in some recent papers that observed 
LFs deviate from the Schechter function due to an excess of very luminous 
galaxies. \citet{G13} found that the H$\alpha$ LF for Galaxy And Mass Assembly (GAMA) 
galaxies with $z<0.35$ is better described by the Saunders function, 
which can be reduced to the log-normal distribution for the considered sample.

The H$\alpha$ and UV radiation is produced by the population of massive stars, which 
were formed during the recent burst. Therefore, galaxy luminosities at these wavelengths 
attain a maximum shortly after a starburst. 
Later on, the H$\alpha$ and UV luminosities strongly decrease with time after a starburst 
age $T_0\approx 3.2$ Myr, when the most massive stars start to fade out. 
The temporal dependence of this rapid luminosity evolution was found by \citet{PII}.
It has a form
\begin{equation}\label{e1}
L(T)=L_0f(T), \,\, f(0)=1,
\end{equation}
where $T$ is the starburst age, $L_0$ is the luminosity at $T=0$, and $f(T)$ is
approximated by
\begin{equation}\label{e9}
f(T)=\left\{
\begin{array}{rl}
1&\textrm{if $T\le T_0$;}\\
\exp(-p(T-T_0))&\textrm{if $T > T_0$,}\\
\end{array}\right.
\end{equation}
with $T_0=3.2$ Myr and the exponent $p$, which depends on the wavelength 
\citep[see details in][]{PII, PI}. The same function is appropriate for the luminosities in the 
22 $\mu$m IR band, derived using the {\sl Wide-field Infrared Survey Explorer} 
({\sl WISE}) data and probably for the radio emission at 1.4 GHz \citep{AN}.

Naturally, the rapid luminosity evolution affects the observed LF. This effect 
was studied by \citet{P}, who derived the functional forms of LFs in the cases 
when the initial galaxy luminosities $L_0$ are approximated by the Schechter 
or log-normal distributions. However, we note that the decrease 
with time of the starburst H$\alpha$ luminosity is much steeper than that
of the FUV and NUV luminosities. This is because the H$\alpha$ emission is
produced only by the most massive short-lived stars while the contribution of 
the less massive and longer-lived stars to the FUV and NUV emission is 
important. 

In particular, between 0 and 10 Myr it decreases by more than two 
orders of magnitude, while the FUV luminosity is decreased only by a factor of 
two \citep{Lei}. \citet{I14} have shown that the FUV-to-H$\beta$ luminosity 
ratio for galaxies with high EW(H$\beta$) is lower than that for galaxies with 
low EW(H$\beta$). This is due to the more rapid evolution of the H$\beta$ 
luminosity. Therefore, H$\alpha$ and FUV luminosities first should be reduced 
to zero age and only after that be compared. 

Our goal is to find functional dependences, which provide the best fitting of 
the sample LFs and to derive sets of their optimal parameters.
We use the methods of the mathematical statistic described e.g. by
\citet{ref:F} and \citet{ref:H}. The maximum likelihood method (MLM) 
is used to derive an optimal set of parameters
and the Pearson's $\chi^2$-test is applied to estimate the accuracy of fitting.

The properties of the sample are described in Section \ref{s:Sample}.
We justify the choice of the approximations for the luminosity functions
in Section \ref{s:LF}. 
Luminosity functions in the H$\alpha$ emission line and UV continuum are 
considered in Sections \ref{s:Ha} and \ref{s:UV}, respectively. A brief
discussion of possible physical mechanisms influencing the LF shape
is presented in Section \ref{s:d}. 
The best LFs approximations are summarised in Section \ref{s:concl}.

\section{Data description}\label{s:Sample}

We use the sample of 795 LCGs selected by \citet*{I11} from the Data Release 7
of the Sloan Digital Sky Survey (SDSS)\citep{A09}.
These data were supplemented by FUV ($\lambda_{\rm eff}$=~1528\AA) 
and NUV ($\lambda_{\rm eff}$ =~2271\AA) fluxes,
extracted by \citet{PII} from the {\sl GALEX} Medium and All-sky Imaging Surveys.

Briefly, the sample selection criteria were as 
follows: high equivalent width EW(H$\beta$) $\geq $ 50\AA\ and high luminosity
$L$(H$\beta$) $\geq $ 3$\times$10$^{40}$ erg s$^{-1}$ of the H$\beta$ emission 
line; well-detected 
[O {\sc iii}] $\lambda $4363 {\AA} emission line in galaxy spectra, with a 
flux error less than 50\%. Only star-forming galaxies with angular 
diameters $\le 10\arcsec$ were selected. All these criteria select 
galaxies with strong emission lines in their spectra at redshifts 
$z$ $\sim$ 0.02 -- 0.6.
The high equivalent width EW(H$\beta$) results in selection of LCGs with
starburst ages $T<5$ Myr.

\citet{I11} showed that LCGs from this sample occupy the region of
star-forming galaxies with high-excitation H {\sc ii} regions 
in the [O {\sc iii}]$\lambda$5007/H$\beta$ vs. 
[N {\sc ii}]$\lambda$6583/H$\alpha$ diagnostic diagram 
\citep[BPT diagram, ][]{BPT}. All LCGs lie below the line \citep{Kauf}
separating star-forming galaxies and active galactic nuclei.

LCGs are characterised by strong and rare bursts of star formation and
their properties are similar to those of so-called ``green pea'' (GP) galaxies
discussed by \citet{C09}. These 
galaxies have low metallicities (of $\sim$ 20 \% solar)
\citep{A10,I11}, low stellar masses of $\sim 10^{8.5}$ -- $10^{10}M_\odot$ 
\citep{C09, I11}, strong [O {\sc iii}] $\lambda\lambda$ 4959, 5007\AA\ emission
lines, high star formation rates \citep[SFR $\sim 10$ M$_\odot$yr$^{-1}$, ][]{C09} and 
high specific star formation rates 
\citep[up to $\sim 10^{-8}$ yr$^{-1}$, ][]{C09,I11}. These features 
place them between nearby blue compact dwarf galaxies on one side and high-redshift 
UV-luminous Lyman-break galaxies and H$\alpha$ emitters on the other side. 

Some parameters of these LCGs including the starburst ages $T$ and the masses of the 
young stellar population $m$ were obtained from fitting their spectral energy distributions
in the wavelength range $\lambda$$\lambda$3800 -- 9200\AA\ 
\citep[see details of selection criteria and derived global characteristics in ][]{I11}.
Neither $T$ nor $m$ were used as selection criteria.

The reddening corrections to H$\alpha$ and 
UV band fluxes are applied adopting the \citet{C89} reddening law. 
The extinction $A(V)$ and reddening $E(B-V)$ of each galaxy
was obtained by \citet{I11} from the hydrogen Balmer decrement in the rest 
frame SDSS spectra. 
Adopting $R(V)$ = $A(V)/E(B-V)$ = 3.1, we obtain
$A$(H$\alpha$) = 2.54$\times$$E(B-V)$, $A$(FUV) = 8.15$\times$$E(B-V)$
and $A$(NUV) = 9.17$\times$$E(B-V)$.
All hydrogen line fluxes were corrected for both the reddening and underlying
stellar absorption.

Note that \citet{I16} checked different reddening laws comparing SDSS optical and 
HST/COS UV spectra of one of the compact galaxies. They found that reddening 
curves with low $R(V)$ $\sim$ 2.7 are more suitable, while the reddening law 
by \citet{Ca94} and \citet{Ca00} is not appropriate.
All H$\alpha$ emission line luminosities were also corrected for aperture.

To verify these corrections we compared the H$\alpha$ line 
luminosities for LCGs with their luminosities in the 
22 $\mu$m IR band, calculated using {\sl WISE} data. 
We find that the H$\alpha$-to-22 $\mu$m band luminosity ratios for our
galaxies are approximately constant, indicating that these luminosities are approximately
proportional with a linear Pearson's correlation coefficient of 90\%. 
Similar proportionality was obtained by \citet*{LHK} 
and \citet{I14} \citep[see also ][]{BC}. Adopting the \citet{C89}
reddening law \citet{I14} found that the FUV-to-H$\alpha$ and NUV-to-H$\alpha$ 
luminosity ratios are also constant. However, the exact values of the latter
ratios depend on the adopted $R(V)$, while H$\alpha$-to-22 $\mu$m band luminosity 
ratios are not sensitive to the value of $R(V)$.

The luminosities were corrected for extinction derived individually in each galaxy. 
According to \citet{SL} only a sample
with individual corrections for extinction can show the deviation of LF 
from the Schechter function. LFs without extinction
correction or corrected using averaged statistical relations have a form that is
similar to the Schechter function by a coincidence.

Summarising, the entire sample consists of 795 galaxies with H$\alpha$ luminosities \citep{PII}.
The majority of these galaxies were also detected in the FUV and NUV ranges.
Additionally we extracted a subsample of 691 galaxies
with a single star-forming region by discarding the galaxies with multiple 
knots of star formation. This subsample is used to study the rapid luminosity
evolution and to compare with theoretical predictions by \citet{P}, which
were made for the galaxies with a single star-forming region.  

\section{Analythical functions for fitting of the LF}\label{s:LF}   

\subsection{Commonly used LF approximations}\label{s:LF1}   

The luminosity function is a very important statistical characteristic 
of galaxy populations. The optical LF can be reproduced by the Schechter function, 
which is also known in the mathematical
statistics as the gamma distribution:
\begin{equation}\label{sc}
\varphi(L)\propto (L/L^*)^{\alpha} \exp(-L/L^*).
\end{equation}
Here $\varphi(L) dL$ is the number of galaxies per unit volume in
the luminosity interval from $L$ to $L+dL$.
Parameters $\alpha$ and $L^*$ are determined from the shape of this function. 
The slope in the [$\ln \varphi(L), \ln L$] plane is equal to $\alpha$ 
for $L\ll L^*$ and is strongly changed at $L\approx L^*$.

We define the probability density of luminosity distribution $n(L)$, called the sample LF, by 
introducing a probability $n(L)dL$ for the galaxy, randomly selected from a
sample, to have the luminosity in 
the interval from $L$ to $L+dL$. If $N$ is the number of galaxies in the sample
then $Nn(L)dL$ is the number of galaxies with the luminosity in 
the interval from $L$ to $L+dL$. 

To find $n(L)$ we multiply the function
$\varphi(L)$ by the volume $V(L)$ occupied by the sample galaxies with the 
luminosity $L$. For a flux-limited sample with small galaxy redshifts we can
adopt $V(L)\propto L^{3/2}$ neglecting the effect of general relativity
and differences between 
the luminosity distance and other types of distances in relativistic cosmology.
Then the sample LF is described by the relation
\begin{equation}\label{sc1}
n(L)=\frac{u^{\alpha+3/2}e^{-u}}{L^*\Gamma(\alpha+5/2)},\quad u=\frac{L}{L^*}.
\end{equation}
Here $\Gamma$ is the gamma-function. The constant $1/\Gamma(\alpha+5/2)$ is derived from the 
normalisation condition
\begin{equation}\label{e8}
\int_0^{\infty} n(L)dL=1.
\end{equation}

The deviation of LFs from the Schechter function was noted for the far-IR (FIR) 
luminosities at 60 $\mu$m \citep*{Law,Sau,T} or radio luminosities 
\citep*{Con, Wi, Mach}. 
On the other hand, until recently, it was suggested that
LFs for the H$\alpha$ line \citep{Gal,Ly} and 
the UV emission \citep{W} can well be fitted by a Schechter function.

At the same time all these luminosities are indicators of star formation \citep{Ca12}, 
therefore their extinction-corrected LFs must be similar. 
Since FIR and radio LFs are not described by the Schechter function,
the deviation of the H$\alpha$ and UV LFs from this function
is also expected. However, this deviation was found only recently. A reason is that
the luminosities in H$\alpha$ line and UV ranges must be individually corrected for the 
dust extinction, unlike the FIR or radio luminosities. Without this correction
the deviation of LF from the Schechter one is masked.

\citet{SL} showed that the Schechter function can not reproduce   
the distributions of SFRs derived from the UV and H$\alpha$ fluxes,
and therefore their LFs are also not fitted by the Schechter function. On the other hand, 
optical and near-infrared LFs are well reproduced by the Schechter function.
This conclusion was made on the basis of the
UV SFR-to-mass ratio distribution of about 50,000 galaxies in the local 
universe with $z\le 0.1$ \citep{Sal}.

\citet{j13} studied the UV LF of galaxies from the {\sl Galaxy Evolution Explorer} ({\sl GALEX})
Medium Imaging Survey with spectroscopic redshifts from the DR1
of the WiggleZ Dark Energy Survey \citep{D10}. They found an excess of most luminous 
galaxies compared to the Schechter distribution at redshifts greater than $z = 0.55$.

\citet{BD} claimed that the bright end of the UV LF is declined not as steeply as that
predicted by the 
Schechter function of fainter galaxies for the sample of 34 luminous 
galaxies in the redshift range $6.5 < z < 7.5$ from the UltraVISTA DR2 and 
UKIDSS UDS DR10 surveys.  

Similarly \citet*{PII} 
found the excess of galaxies at the bright end of the H$\alpha$, FUV and NUV LFs
for LCGs.
In this paper we study LFs of this sample in more detail.

\citet{G13} found that the H$\alpha$ LF for GAMA 
galaxies with $z<0.35$ is not well described by the Schechter function. 
The exponential drop of the Schechter function at the bright end is too 
steep to match the observed LFs, which are better approximated by the Saunders 
function \citep{Sau}
\begin{equation}\label{saund}
\begin{array}{c}
n(L)=\textrm{const}\times u^{\beta}\exp\left[-A\log^2(1+u)\right],\\
u=L/L^*_s,\quad A=1/(2\sigma^2).
\end{array}
\end{equation}
Here $\sigma$ and $L^*_s$ are constants. Note that \citet{Sau} 
used the notation $L^*$ in (\ref{saund}). We add the subscript ``$s$'' to distinguish this 
quantity from the $L^*$ in (\ref{sc}). Fitting LF of our sample we find that the exact 
value of $L^*_s$ does not change an accuracy of a fit. Since the luminosities of 
the sample galaxies $L_i$ are much greater than $L^*_s$, the unity in the
$(1+u)$ term can be dropped and Eq. (\ref{saund}) is transformed
to the two-parametric log-normal distribution
\begin{align}
\label{en1}
&n(L)=\left(\frac{a}{\pi}\right)^{1/2}\exp\left(-\frac{1}{4a}\right)\, \tilde{L}^{-1}\exp\left(-a\ln^2(L/\tilde{L})\right) \nonumber\\
&=\left(\frac{a}{\pi}\right)^{1/2}\exp\left(-\frac{1}{4a}\right)\,\tilde{L}^{-1}\left(\frac{L}{\tilde{L}}\right)^{-a\ln\left(L/\tilde{L}\right)}.
\end{align}
Here
\begin{equation}\label{en2}
a=A(\ln 10)^{-2},\quad \ln \tilde{L}=\ln L^*_s+\beta/(2a).
\end{equation}
 The set of three parameters ($L^*_s,A,\beta$) can be reduced to the the set of two 
parameters ($\tilde{L},a$). We note that the intrinsic 
parameter $\tilde{L}$ is defined by the combination of $L^*_s$ and $\beta$
only if $L_i\gg L^*_s$.

\subsection{An effect of a rapid luminosity evolution}\label{s:LF2}   

\citet{P} studied the impact of the rapid luminosity evolution on the LF of a
sample of star-forming galaxies by introducing the initial, current, time-averaged and 
sample LFs. He derived the relations between different types of LFs for the case of an 
arbitrary luminosity evolution defined by equation (\ref{e1}) \citep{P}. Considering
flat space and neglecting the differences between
the luminosity distance and other types of distances
\citet{P} obtained the sample LF in the form
\begin{equation}\label{eq1}
n(L)=C\left[ n_1(L)+qn_2(L)\right],
\end{equation}
where the constant $C$ is derived from the normalisation condition (\ref{e8}). 
Here $n_1(L)$ is a sample LF if the rapid luminosity evolution is neglected, 
e.g. in the case of constant galaxy luminosities and 
$n_2(L)=L^{1/2}\int_L^{\infty}x^{-3/2}n_1(x)dx$ is the term that takes into account rapid
luminosity evolution. The constant $q$ is the combination of the parameters from (\ref{e9})
\begin{equation}\label{e11}
q=(T_0p)^{-1}.
\end{equation}
It can be considered as a free parameter and can be obtained from the best fit of the sample
LF. In any case, the condition $q\geq 0$ should be satisfied. 
The case with $q=0$ corresponds to $n(L)=n_1(L)$, i.e to the LF 
without the luminosity evolution.

\citet{P} obtained the sample LFs for the cases with initial galaxy luminosities and $n_1(L)$,
which are approximated by the Schechter function (\ref{sc1}) and the log-normal 
distribution (\ref{en1}).
In the first case
\begin{equation}\label{e15}
n(L)=\frac{u^{\alpha+3/2}e^{-u}+qu^{1/2} \Gamma(\alpha+1,u)}{L^{*}(1+2q/3)\Gamma(\alpha+5/2)}, u=\frac{L}{L^*}.
\end{equation}
This relation includes the incomplete gamma-function, \citep[see e.g. ][]{BE}
\begin{equation}\label{e13}
\Gamma(\alpha+1,u)=\int_u^\infty x^{\alpha} e^{-x} dx
\end{equation}
and depends on three parameters $q$, $\alpha$ and $L^*$.

In the second case
\begin{align}\label{enn}
n(L)=&\tilde{L}^{-1}\left(1+\frac{2q}{3}\right)^{-1}\times \nonumber\\
&\times\left[\left(\frac{a}{\pi}\right)^{1/2}\exp\left(-\frac{1}{4a}\right)\exp\left(-a\ln^2(L/\tilde{L})\right)+\right. \nonumber\\
&+\frac{q}{2}\exp\left(-\frac{3}{16a}\right)\left(\frac{L}{\tilde{L}}\right)^{1/2}\times \nonumber\\
&\left.\times\textrm{erfc}\left(\sqrt{a}\ln\left(\frac{L}{\tilde{L}}\right)+
\frac{1}{4\sqrt{a}}\right)\right].
\end{align}
Here
\begin{equation}\label{enn1}
\textrm{erfc}(x)=1-\textrm{erf}(x)=\frac{2}{\sqrt{\pi}}\int_x^{\infty}e^{-t^2}dt.
\end{equation}
It also depends on three parameters $q$, $a$ and $\tilde{L}$.

The relations (\ref{e15}) and (\ref{enn}) at $q=0$ are transformed to the Schechter function 
(\ref{sc1}) and the log-normal distribution (\ref{en1}), respectively. 
The properties of relations (\ref{e15}) and (\ref{enn}) were analyzed by \citet{P}.

\section{Approximation of the H$\alpha$ LF for the LCG sample}\label{s:Ha}

Using MLM and functions (\ref{sc1}), (\ref{en1}), (\ref{saund}), (\ref{e15}) and (\ref{enn})
we derive optimal sets of parameters of the H$\alpha$ LF for the sample of LCGs.
The accuracy of the fits is verified with the Pearson's $\chi^2$-test. 

\subsection{Fitting the LF by Schechter function}\label{s:comp}

We find a set of parameters 
$\alpha, L^*$, for which the function
\begin{equation}\label{e17}
U=\sum_{i=1}^N \log(n(L_i| \alpha, L^*))
\end{equation}
with $n(L_i| \alpha, L^*)$ from Eq. (\ref{sc1}) attains a maximum. Here
$L_i$ are the extinction- and aperture-corrected luminosities of individual galaxies from
the entire sample.
\begin{figure}
\includegraphics[width=\columnwidth]{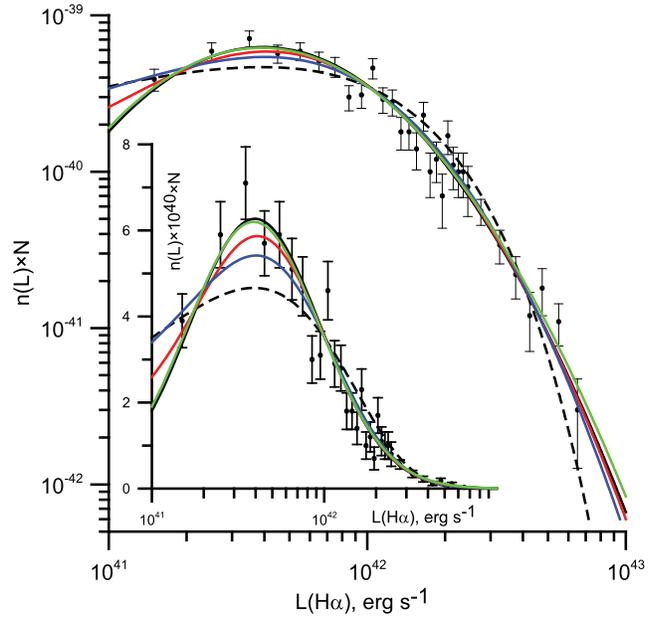}
\caption{The H$\alpha$ LF of the entire LCG sample and its approximation by the 
log-normal distribution (\ref{en1}) (solid black line) and Schechter function 
(\ref{sc1}) (dashed line). Red, blue and green lines show the functions (\ref{enn},\ref{e11}), (\ref{enni})
and (\ref{saund}) with optimal parameters.  
Error bars correspond to the Poisson distribution. The smaller inner frame shows the same plot with linear ordinate axis
to emphasize the differences between the approximations near their maximum.
The differences at the high-luminosity end are best seen in the main plot}
\label{f0}
\end{figure}

The maximum of Eq. (\ref{e17}) is attained if
\begin{equation}\label{e19}
\ln(2.5+\alpha) -\psi(2.5+\alpha)=\ln\left<L\right> -\left<\ln L\right>,
\end{equation}
\begin{equation}\label{e20}
L^*=\frac{\left<L\right>}{2.5+\alpha}.
\end{equation}
Here $\psi$ is the digamma function or the logarithmic derivative
of the gamma function, and angle brackets mean the averaging over the sample. 
These maximum conditions are well known and were used e.g. 
by \citet{KKKP} and \citet*{EEP}.

For the entire LCG sample from (\ref{e19}, \ref{e20}) we derive $\alpha=-1.04$, 
$L^*=8.5\times 10^{41}$ erg s$^{-1}$. The Schechter distribution with these parameters 
predicts the mean number of galaxies entering 
in 9 bins with width $5\times 10^{41}$ erg s$^{-1}$ to be 203, 205, 145, 94.5, 59, 36, 
21.6, 12.8, 7.5, respectively, and 10.4 galaxies with luminosities greater than 
$4.5\times 10^{42}$ erg s$^{-1}$. 
The respective number of galaxies from the observed LF is 227, 218, 139, 66, 56, 28, 17, 
11, 6, 27. The excess of the most 
luminous galaxies is obvious. The Schechter approximation is shown in Fig. \ref{f0} by 
a dashed line. The similar LF excess is present for the galaxies with a single knot of 
star formation approximated by the Schechter function with parameters 
$\alpha=-0.96$, $L^*=8.0\times 10^{41}$ erg s$^{-1}$. We derive
$\chi^2$ values of 42.2 and 38.9 for the entire sample and the subsample of galaxies 
with a single knot of star formation, respectively. 
The probabilities that LFs for these samples are reproduced by the Schechter function are 
$1.7\times 10^{-7}$ and $7.5\times 10^{-7}$ i.e. negligible.

According to \citet{SL}, the individual corrections for extinction are necessary to establish 
the deviation of LF from the Schechter function. The similar conclusions were drawn by 
\citet{G13}. To verify the effect of the correction for extinction we consider the 
``uncorrected'' H$\alpha$ LF for LCGs without any correction. This LF is
approximated by the Schechter function with parameters $\alpha=-1.08$, 
$L^*=3.4\times10^{41}$ erg s$^{-1}$. The predicted mean number and actual number
(in parentheses) entering in 8 bins with width 
$2.5\times 10^{41}$ erg s$^{-1}$ are 270(307), 228(224), 137(123), 76(54), 41(34), 21(24), 
11(8), 5.5(9), respectively, and 5.6(12) galaxies with luminosities greater than 
$2\times 10^{42}$ erg s$^{-1}$. The corresponding
value of $\chi^2$ is 24.4, which is roughly a half of the $\chi^2$ value for the sample 
with the luminosities, corrected for extinction. 

It is clear that the correction for extinction with a mean extinction for all galaxies would
result in the $L^*$ increased by the corresponding factor, while the values of $\alpha$ and 
$\chi^2$ remain the same. Thus, the individual corrections for extinction considerably 
increase the $\chi^2$ value in agreement with the statement of \citet{SL}. 
Nevertheless, the value of $\chi^2$ for the deviation of the ``uncorrected'' LF from the 
Schechter function is big enough. For 6 degrees of freedom it corresponds to the probability 
$\approx 0.04\%$ that LF is defined by the Schechter distribution.
This difference arises because of an excess of galaxies on the bright end of LF and a deficit 
of galaxies in the middle region 
$7.5\times 10^{41}$ erg s$^{-1}<L<10^{42}$ erg s$^{-1}$.

Thus, we can reject the possibility that sample LF is described by the Schechter function. 
Moreover, the LF deviates from the Schechter function even without individual luminosity 
correction. 

\subsection{Fitting the LF by log-normal function}\label{s:ln}

The MLM optimal parameters for the distribution (\ref{en1}) are
\begin{equation}\label{en3}
\begin{array} {c}
\ln \tilde{L}=\langle\ln L\rangle +\langle\ln L\rangle^2-\langle\ln^2 L\rangle,\\ 
a=\left[2\left(\langle\ln^2 L\rangle-\langle\ln L\rangle^2\right)\right]^{-1}.
\end{array}
\end{equation}
Angle brackets mean the averaging over the sample. 

The log-normal distribution is more intuitive, it has no unnecessary parameters and its
parameters can be easily obtained from Eq. (\ref{en3}). The distribution (\ref{en1})
is normalized according to (\ref{e8}). Note that the distribution of the
probability density over $\ln L$
achieves the maximum at $\tilde{L}_1=\exp\langle\ln L\rangle$, not at $\tilde{L}$.

For the H$\alpha$ LF of the entire LCG sample we derive
$a=0.657$, $\tilde{L}=3.96\times 10^{41}$ erg s$^{-1}$ and $\tilde{L}_1=8.5\times 10^{41}$ 
erg s$^{-1}$. This approximation is shown in 
Fig. \ref{f0} by solid black line. The $\chi^2$-test provides the probability 27\% that
deviations are random with the value $\chi^2=8.7$ and 7 degrees of freedom (d.o.f.).
Similarly, for the subsample with a single knot of the star formation we obtain a fit with
$a=0.689$, $\tilde{L}=4.12\times 10^{41}$ erg s$^{-1}$ and 
$\tilde{L}_1=8.5\times 10^{41}$ erg s$^{-1}$. Its $\chi^2=10.2$ corresponds
to 18\% probability that the difference in LFs is due to the random deviations.

\subsection{Fitting the LF by Saunders function}\label{s:Sau}

Calculating the optimal set of parameters for Saunders function (\ref{saund}) using 
MLM, we meet with the asymptotic degeneration, when the different sets of $\beta$,
$L^*$ and $A$ correspond to the distribution (\ref{en1}) with the same
parameters $a$ and $\tilde{L}$. Consequently, the minimal $\chi^2$ for the fit
of LF by the Saunders function should be less or equal to the value obtained by
the log-normal fit. 

Numerical calculations for the entire LCG sample show that a minimal 
value $\chi^2=7.1$ is achieved at 
$L^*_s=8\times 10^{39}$ erg s$^{-1}$, $\beta=4.69$
and $A=3.25$. The function (\ref{saund}) with this set of parameters is shown in Fig. \ref{f0} by the green line.
The number of d.o.f. for $\chi^2$ is 6 for the three-parameter Saunders function, therefore
the minimal value $\chi^2=7.1$ corresponds to the probability of 31\% that the deviations 
between LF and the Saunders function (\ref{saund}) are due to the random errors.
This means that the fit by the Saunders function is as accurate as the fit by 
the log-normal distribution. We will show in section \ref{s:lnq} that the obtained 
probability is less than that for the fit by function (\ref{enn}).

For the subsample with a single knot of star formation the minimal value $\chi^2=9.0$ is 
attained with the set
$L^*_s=8\times 10^{39}$ erg s$^{-1}$, $\beta=5.01$ and $A=3.43$. 
It corresponds to the probability of 17\%
that the difference between LF and the Saunders function is random.

Thus, we confirm the conclusion of \citet{G13} that the Saunders function (\ref{saund}) 
is much better approximation of LF than the Schechter one. The bright end with 
$L>2.5\times 10^{42}$ erg s$^{-1}$ is fitted very well.
We obtained nearly the same result with the log-normal function.
The main contribution to the $\chi^2$ value is provided by the deficit of galaxies in the 
$2\times 10^{42}$ erg s$^{-1}<L<2.5\times 10^{42}$ erg s$^{-1}$ bin. 

\subsection{LF fitting by the Schechter function in the case of rapid luminosity
evolution}\label{s:aq}

\begin{figure}
\includegraphics[width=\columnwidth]{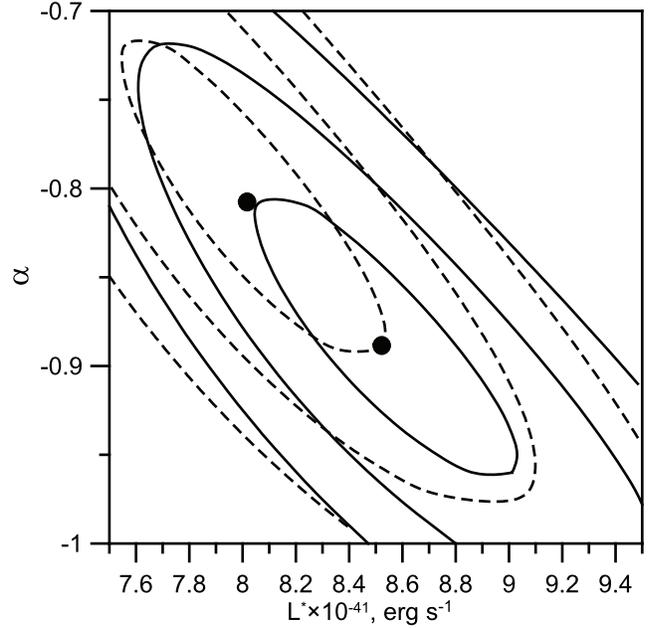}
\caption{The borders of 1$\sigma$-, 2$\sigma$- and 3$\sigma$- confidence 
regions according to MLM. 
Solid and dashed lines correspond to the entire sample and the subsample of 
LCGs with a single knot of the star formation, respectively}
\label{f1}
\end{figure}

We approximate the luminosity function of the sample by (\ref{e15}) and 
find optimal values of its parameters using MLM. \citet{PII} derived $p=0.65$ 
Myr$^{-1}$ and $T_0=3.2$ Myr
for the entire sample, therefore we obtain $q=0.48$ from (\ref{e11}). 
Adopting this value, we obtain the optimal MLM values
$L^*=(8.5\pm 0.5) \times 10^{41}$ erg s$^{-1}$ and $\alpha=-0.88\pm 0.07$ 
at the 68\% confidence level. 
In Fig. \ref{f1} we plot by solid lines the borders of $1\sigma$, $2\sigma$ and $3\sigma$ 
confidence regions of $\alpha$ and $L^*$. 
By dashed lines we also plot the borders of the $1\sigma$, 
$2\sigma$ and $3\sigma$ confidence regions for the subsample 
with a single knot of the star formation.
The optimal values  of parameters for this subsample are 
$L^*=(8.0\pm 0.5) \times 10^{41}$ erg s$^{-1}$ and $\alpha=-0.81\pm 0.07$. 
We find that the $1\sigma$ confidence regions for the entire sample and the subsample 
are substantially overlapped.

Using the Pearson's $\chi^2$ test we obtain $\chi^2=37.4$ and 33.3
with 7 d.o.f. for the entire sample and the subsample with a 
single knot of star formation, respectively. These values correspond
to the probability less than 0.01\% for the luminosity function of the samples 
to be of the form of Eq. (\ref{e15}). 

The derived parameter $L^*$ is the same as that obtained in Section
\ref{s:comp} by fitting with the Schechter function (\ref{sc1}) and neglecting the
rapid luminosity evolution. On the other hand, the values of $\alpha$ in the case 
of the luminosity evolution
are greater by $2.3\sigma$ for the entire sample
and by $2.1\sigma$ for the subsample of galaxies with a single knot of 
star formation than the values obtained with the Schechter function 
Eq. (\ref{sc1}) . Thus, $\alpha$ is considerably 
underestimated if the rapid luminosity evolution is neglected. 

The above conclusions are obtained for the fixed value of $q$. Now we consider
the case with the varying $q$.
There is the problem preventing MLM from finding the unique set of three optimal 
parameters considered by \citet{P}. Two completely different sets of parameters are
resulted in practically the same shapes of function (\ref{e15}). One can see
in Fig. 1 from \citet{P} that the function with
$\alpha=-0.51$, $L^*=8.08\times10^{41}$ erg s$^{-1}$, $q=2$ is almost coincident with the
function (\ref{e15}) and the set of parameters obtained above, namely
$\alpha=-0.88$, $L^*=8.5\times10^{41}$ erg s$^{-1}$, $q=0.48$. 
Consequently, the function $U$ from (\ref{e17}) has two equal maxima and a prolate form 
of confidence interval.

Nevertheless, varying $q$ and using MLM we search for two
other parameters. For the entire sample we obtain the MLM values
$\alpha=-0.77, -0.69, -0.63, -0.42, -0.39$, $L^*=(8.57, 8.59,$ $8.63, 9.16, 
9.70)\times 10^{41}$ erg s$^{-1}$ and $\chi^2=34.0, 32.7, 31.7,$ $27.8, 27.2$ 
for $q=1, 1.5, 2, 10, 100$, 
respectively. The plots of all distributions with these sets of parameters 
are very similar, while parameters $\alpha$ and $L^*$ increase with 
increasing $q$. The maximum on the plot becomes 
slightly higher and narrower with increasing $q$. Consequently, $\chi^2$ is
slightly decreased. The minimum of $\chi^2$ is attained at $q\to \infty$. 
The likelihood for the subsample with a single knot of star formation is 
maximal at $q\to \infty$ with
$\alpha=-0.26$, $L^*=9.1\times 10^{41}$ erg s$^{-1}$, and $\chi^2=25.2$,
corresponding to 0.14\% probability that deviations of the fit from the LF are statistical. 

Thus, Eq. (\ref{e15}) is a very poor approximation of 
the luminosity function of LCGs, mainly due to the excess of very luminous 
galaxies.

\subsection{LF fitting by the log-normal function in the case of rapid luminosity
evolution}\label{s:lnq}

To fit the LF of the entire LCG sample with the MLM we use the function (\ref{enn})
adopting $q=0.48$ and find $a=0.73$ and 
$\tilde{L}=4.9\times 10^{41}$ erg s$^{-1}$. 
The function (\ref{enn}) with this set of parameters is 
shown in Fig. \ref{f0} by the red line. The derived $\chi^2=7.6$ corresponds
to 37\% probability that the difference in LFs is caused by random deviations.
For the subsample with a single knot of star formation we find 
$a=0.77$, $\tilde{L}=5.1\times 10^{41}$ erg s$^{-1}$ and $\chi^2=9$,
corresponding to the probability of 25\%.

Varying $q$ we can
find the dependence of the MLM function $U$ and $\chi^2$ on $q$. 
For the subsample with a single knot of star formation $U$ decreases if $q$ increases.
However, this dependence is very weak.

We give some values of $a$ and $\tilde{L}$ for different $q$.  
Adopting $q=0.05,0.1,1,2,5,10,30,100$ we find
the values 0.70, 0.71, 0.83, 0.90, 0.98, 1.01, 1.02, 1.02 for $a$ and 
4.25, 4.36, 5.87, 6.85, 8.16, 8.9, 9.4, 9.6 erg s$^{-1}$ for $\tilde{L}\times 10^{-41}$, respectively.
The respective values of $\chi^2$ are 10.4, 9.9, 8.6, 8.5, 8.7, 8.7, 8.8, 8.8
with the minimum at $q\approx 2$. Using this value we decrease the d.o.f. by 1. 
The minimal value $\chi^2=8.5$ with 6 d.o.f. corresponds to the 20\% probability 
that the difference between LFs is random. This probability is worse than the value
obtained above with $q=0.48$ according to (\ref{e11}) with 7 d.o.f.
 
Similar conclusions can be drawn for the entire sample. 
Adopting $q=0.05,0.1,0.25,1,2,6,10,50,1000$ 
we find the values 0.668, 0.675, 0.702, 0.789, 0.854, 0.919, 0.943, 0.958, 0.959 for $a$ 
and 
4.09, 4.18, 4.49, 5.66, 6.62, 7.99, 8.51, 9.19, 9.36 erg s$^{-1}$ for $\tilde{L}\times 10^{-41}$, respectively. The value of $U$ slightly decreases if $q$ increases.
The respective values of $\chi^2$ are 8.65, 8.4, 8.3, 
7.5, 7.5, 7.3, 7.6, 7.7, 7.7 with the minimum at $q\approx 2.5$. 
The difference between
minimal $\chi^2$ and its value at $q=0.48$ according to (\ref{e11}) is not 
sufficient to compensate the reduction of d.o.f. number in attemps to increase the probability.

Summarising, there are three special values of $q$ which do not reduce
the number of d.o.f. These are the limiting values $q=0$ and $q\to \infty$ and the 
condition (\ref{e11}), which is not associated with minimal $\chi^2$ value. 
We find that the function (\ref{enn}) with $q=0.48$ according to (\ref{e11}) 
provides a better fit compared to log-normal one. 

The asymptotics of the function (\ref{enn}) at $q \to \infty$ has the form
\begin{align}\label{enni}
n(L)&=\frac{3}{4\tilde{L}}\exp\left(-\frac{3}{16a}\right)\left(\frac{L}{\tilde{L}}\right)^{1/2}\nonumber\\
&\times\textrm{erfc}\left(\sqrt{a}\ln\left(\frac{L}{\tilde{L}}\right)+
\frac{1}{4\sqrt{a}}\right) \nonumber\\
&=\frac{3}{4L_{c}}\exp\left(-\frac{9}{16a}\right)\left(\frac{L}{L_{c}}\right)^{1/2}\nonumber\\
&\times\textrm{erfc}\left(\sqrt{a}\ln\left(\frac{L}{L_{c}}\right)\right),
\end{align}
where $L_{c}=\tilde{L}\exp(-1/4a)$. The function (\ref{enni}) has two parameters.
According to MLM the best fit of the H$\alpha$ LF for the entire sample is achieved with
$a=0.96$, $L_{c}=7.23\times 10^{41}$ erg s$^{-1}$ and 
$\tilde{L}=9.4\times 10^{41}$ erg s$^{-1}$. The function (\ref{enni}) with this set of parameters is 
shown in Fig. \ref{f0} by the blue line. The value $\chi^2=7.73$ and the 36\% probability that the 
difference in LFs is caused by random deviations are very similar to the values
obtained for the function  
(\ref{enn}) and the value of $q=0.48$ according to (\ref{e11}). The same is true also for the subsample of LCGs 
with a single knot of star formation. The best approximation is achieved with $a=1.02$, 
$L_{c}=7.6\times 10^{41}$ erg s$^{-1}$ and $\tilde{L}=9.7\times 10^{41}$ erg s$^{-1}$. 
The value $\chi^2=8.85$ corresponds to the probability of 26\%.

The LF fits by functions (\ref{enn},\ref{e11}) and (\ref{enni}) have similar accuracy
and are better than the fit by the log-normal function.
\begin{figure}
\includegraphics[width=\columnwidth]{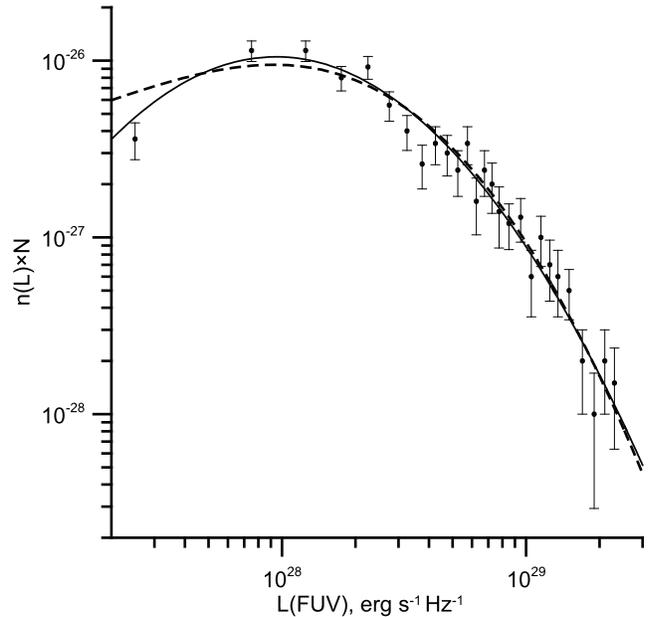}
\caption{The FUV LF of the entire LCG sample and its approximation by the log-normal
distribution (\ref{en1}) (solid line) and function (\ref{enni})
(dashed line). 
Error bars correspond to the Poisson distribution}
\label{FUV}
\end{figure}
\begin{figure}
\includegraphics[width=\columnwidth]{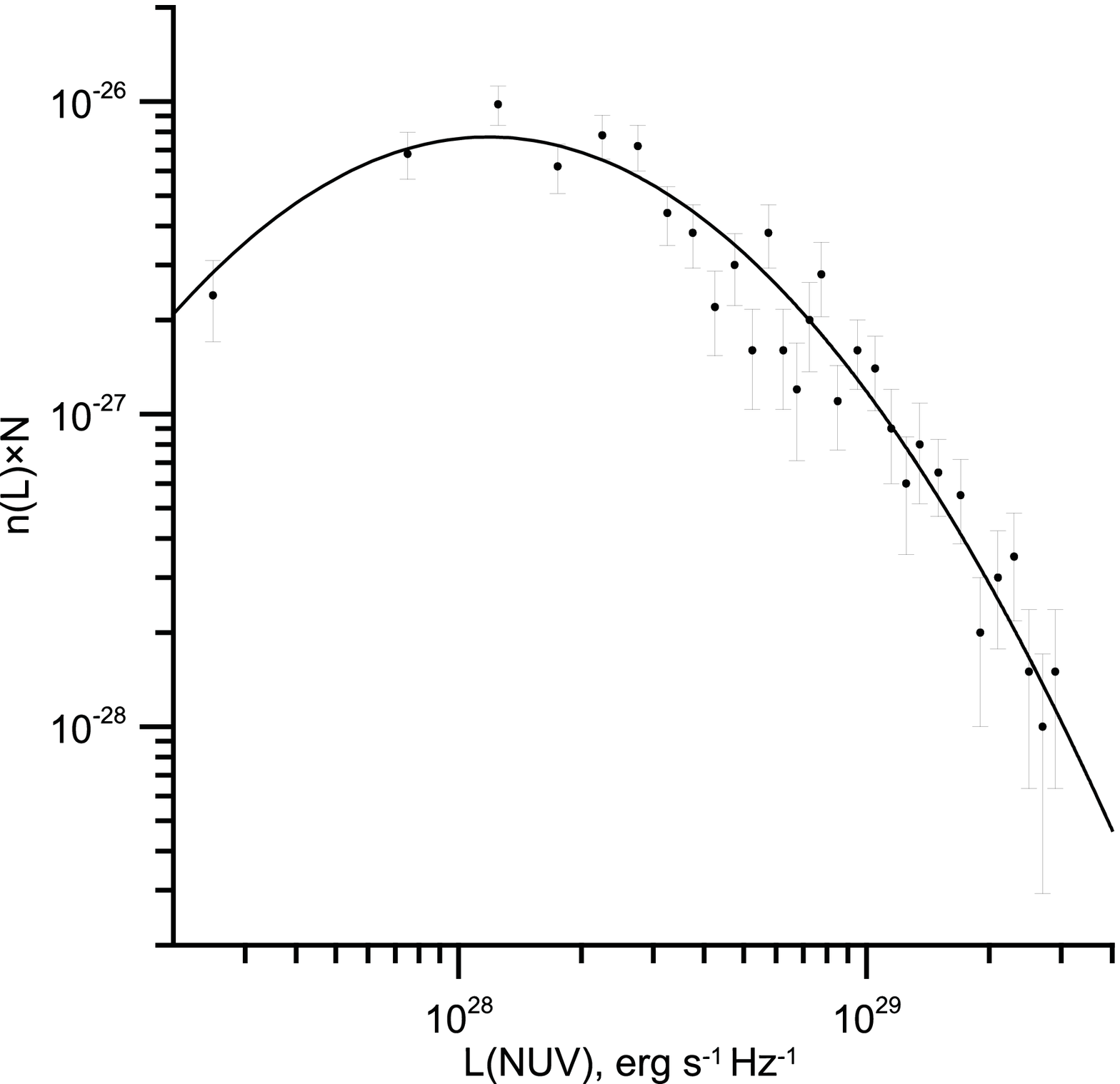}
\caption{The NUV LF of the entire LCG sample and its approximation by the log-normal
distribution (\ref{en1}). Error bars correspond to the Poisson distribution}
\label{NUV}
\end{figure}

\subsubsection{A generalization of function (\ref{enni})}\label{s:erfci}

We found that fits by (\ref{enn}) with $q$ from (\ref{e11}) and (\ref{enni}) are resulted
in similar $\chi^2$ values. The advantage of the function (\ref{enni}) is that it fits the 
high-end of LF very well. However its behaviour at low-end is fixed and is approximated 
by $n(L)\propto L^{1/2}$. On the other hand, the Schechter function badly fits the high-end 
of LF, but its approximation at low-end LF depends on $\alpha$. 
Combining two advantages we obtain the function, which is different from the Saunders 
distribution (\ref{saund}). This function has the form
\begin{align}\label{enng}
n(L)&=\frac{(\gamma+1)^2}{2L_{c}}\exp\left(-\frac{(\gamma+1)^2}{4b^2}\right)\left(\frac{L}{L_{c}}\right)^{\gamma}\nonumber\\
&\times\textrm{erfc}\left(b\ln\left(\frac{L}{L_{c}}\right)\right),\quad b=\sqrt{a}
\end{align}
and it coincides with (\ref{enni}) at $\gamma=1/2$. 
This function is characterised by a very wide maximum and has large errors
of $\gamma$ for our entire LCG sample. 
Nevertheless, it could be used to approximate
the LF of the sample of galaxies with much larger luminosity range as compared to
that for LCG sample. 

\section{Approximation of the UV LFs }\label{s:UV}

To fit FUV and NUV LFs we consider the same functions as those discussed above
for the H$\alpha$ LF. 
The {\sl GALEX} UV luminosities for the galaxies from the entire sample
were collected by \citet{PII} and they are not as accurate as the H$\alpha$ luminosities. 
The sample includes the galaxies with the FUV and NUV luminosity relative 
errors not exceeding 50\%. Using these data we produce the subsamples with the errors 
below 10\%, 20\% and 30\%. We also produce respective subsamples of galaxies 
with a single knot of star formation.

To study the effect of measurement errors, we compare the results
varying the threshold of luminosity errors. On this basis we identify the best subsample.
We select the distribution Eq. (\ref{e15}) with condition Eq. (\ref{e11}) and apply MLM
to find the parameters $\alpha$ and $L^*$. We show in Table \ref{t1} the 
results of computations for all subsamples. 
The values of $p$ are taken from \citet{PI}. They are used for calculation 
of $q$ from Eq. (\ref{e11}).

\begin{table*}
\caption{Values of the parameters in Eq. (\ref{e15}) of the UV LFs
for different LCG subsamples. The values of $p$ are taken from \citet{PI}, $q$ values are calculated 
from Eq. (\ref{e11}), $N$ are the numbers of the galaxies in the subsamples. The ``Err'' column indicates 
the threshold of the measured UV flux errors}
\begin{tabular}{l|ccccc|ccccc}
\hline
\multirow{3}{*}{Err}&\multicolumn{5}{c|}{FUV}&\multicolumn{5}{c}{NUV}\\
&\multirow{2}{*}{$N$}&$p$,&\multirow{2}{*}{$q$}&\multirow{2}{*}{$\alpha$}&$\bigstrut L^*\times 10^{-28},$&
\multirow{2}{*}{$N$}&$p$,&\multirow{2}{*}{$q$}&\multirow{2}{*}{$\alpha$}&$\bigstrut L^*\times 10^{-28}$,\\
&&Myr$^{-1}$&&&erg s$^{-1} \textrm{Hz}^{-1}$&&Myr$^{-1}$&&&erg s$^{-1} \textrm{Hz}^{-1}$\\
\hline
\multicolumn{11}{c}{All galaxies} \\ 

10\%&215&0.43&0.73&$-1.26\pm0.12$&$4.6\pm 0.5$&214&0.33&0.95&$-1.40\pm0.10$&$\bigstrut 7.9^{+1.0}_{-0.8}$\\
20\%&463&0.35&0.89&$-1.29\pm0.08$&$4.8\pm 0.4$&462&0.30&1.04&$-1.37\pm0.07$&$7.6\pm 0.6$\\
30\%&557&0.43&0.73&$-1.34\pm0.06$&$5.1\pm 0.4$&556&0.33&0.95&$-1.43\pm0.06$&$8.3\pm 0.6$\\
50\%&630&0.43&0.73&$-1.33\pm0.05$&$4.9\pm 0.3$&667&0.33&0.95&$-1.44\pm0.05$&$8.3\pm 0.6$\\
\multicolumn{11}{c}{With single knot of starformation} \\ 

10\%&172&0.51&0.61&$-1.26\pm0.12$&$4.3\pm 0.6$&172&0.39&0.80&$-1.41\pm0.11$&$\bigstrut 7.4^{+1.1}_{-0.9}$\\
20\%&393&0.39&0.80&$-1.29\pm0.08$&$4.8\pm 0.4$&393&0.34&0.92&$-1.38\pm0.08$&$7.4\pm 0.7$\\       
30\%&481&0.46&0.68&$-1.32\pm0.07$&$4.9\pm 0.4$&481&0.36&0.87&$-1.42\pm0.07$&$7.9\pm 0.6$\\       
50\%&545&0.47&0.66&$-1.31\pm0.06$&$4.6\pm 0.4$&583&0.37&0.84&$-1.43\pm0.06$&$7.8\pm 0.6$\\  \hline      
\end{tabular}\label{t1}
\end{table*}

\begin{table*}
\caption{Parameters of different fits of FUV and NUV LFs for subsamples with the
20\% accuracy threshold. The optimal values of parameters $a$ and $\tilde{L}$ or $L_{c}$ 
are derived with the MLM. The values $q$ from
(\ref{e11}) are used in functions (\ref{enn}). We denote by $p(H_0)$ the probability that 
null hypothesys is correct and all LF deviations from the corresponding approximation are 
random. The $p(H_0)$ values are calculated from the $\chi^2$ values with 6 d.o.f.}   
\begin{tabular}{@{}l|ccccc|ccccc@{}}
\hline
&\multicolumn{5}{c|}{FUV}&\multicolumn{5}{c}{NUV}\\
Fit&$a$&$\bigstrut L$, erg s$^{-1} \textrm{Hz}^{-1}$&$q$&$\chi^2$&$p(H_0),\%$&
$a$&$\bigstrut L$, erg s$^{-1} \textrm{Hz}^{-1}$&$q$&$\chi^2$&$p(H_0),\%$\\
\hline
(\ref{en1})&0.445&$\bigstrut\tilde{L}=9.46\times 10^{27}$&-&4.48&61&0.412&$\tilde{L}=1.19\times 10^{28}$&-&5.34&50\\
(\ref{enn},\ref{e11})&0.524&$\bigstrut\tilde{L}=1.41\times 10^{28}$&0.89&3.77&70&0.48&$\tilde{L}=1.81\times 10^{28}$&1.04&5.77&45\\
(\ref{enni})&0.609&$\bigstrut L_{c}=1.61\times 10^{28}$&-&1.75&94&0.544&$L_{c}=1.91\times 10^{28}$&-&6.76&34\\
\hline      
\end{tabular}\label{t2}
\end{table*}
 
Results for different subsamples are not very different.
There is a weak negative trend of values with the adopted value of the threshold.
The $\alpha$ values are approximately the same for all subsamples, while the $L^*$ 
values are slightly smaller for subsamples with single star-forming knot. 

The parameter errors for subsamples with 
10\% accuracy threshold are larger due to a small number of galaxies. 
Lower parameter errors are obtained for subsamples 
with 20\% or 30\% accuracy thresholds. For clarity, we select the subsamples with 
20\% accuracy threshold as reference ones.

Applying the Schechter function (\ref{sc1}) for subsamples with single knots of star formation
and 20\% accuracy threshold we obtain $\alpha=-1.43\pm 0.06$, 
$L^*=(4.7\pm 0.4) \times 10^{28}$ erg s$^{-1} \textrm{Hz}^{-1}$ and $\alpha=-1.52\pm 0.06$, 
$L^*=(7.3\pm 0.6) \times 10^{28}$ erg s$^{-1} \textrm{Hz}^{-1}$ for FUV and NUV ranges, 
respectively. Note that $\alpha$ values are lower by 0.14 than the values 
in Table \ref{t1}, which were calculated with $q$ from (\ref{e11}). 
The $\chi^2$ test shows the statistically significant excess of the most 
luminous galaxies as compared to the distributions (\ref{sc1}) and (\ref{e15}). 
We conclude that both the initial and the sample UV LFs of these samples differ 
from the Schechter function at more than 99\% confidence level.

We calculate the parameters for the reference subsamples using the 
log-normal function (\ref{en1}), functions (\ref{enn},\ref{e11}) and (\ref{enni}) and show 
them in Table \ref{t2}. We also apply MLM adopting (\ref{enn}) and different $q$ values.
The confidence region of MLM is very prolate and the dependence $U(q)$ on optimal 
parameters $a(q)$, $\tilde{L}(q)$ is very weak. In this case MLM does not allow
to derive the best $q$. Therefore, we directly use the function $\chi^2(q)$. 
It has a minimum at $q\to \infty$ for the FUV LF and at $q=0$ for the NUV LF.
Thus, we certainly can restrict ourselves to the functions (\ref{en1}) with $q=0$, (\ref{enn}) 
with $q$ from (\ref{e11}) and (\ref{enni}) with $q\to \infty$. 
All other values of $q$ correspond to the parameters 
obtained from the dataset with the number of d.o.f., which is decreased by 1, and
all $\chi^2(q)$ minima are coincident with those for one of three special cases.

\section{Discussion}\label{s:d}

We discuss here what physical mechanisms make LFs of LCGs 
to be different from the LF in the 
optical continuum of the quiescent galaxies, which can be fitted by the
Schechter function. 

Our LCG sample consists of the galaxies with the high star-formation activity
and starburst ages less than 5 Myr. This gives a unique possibility to study 
the LF of the young stellar population  just after the onset of star formation. 
The sample was constructed by \citet{I11} to minimize uncertainties introduced 
by some effects. Small angular diameters minimize errors for aperture 
corrections. The diagnostic diagram was used to select only star-forming 
galaxies. Criterion on strong nebular [O {\sc iii}] emission lines selects 
galaxies with high ionization parameters and therefore minimizes uncertainties 
in the element abundances with the use of strong-line methods. Low limit on 
H$\beta$ luminosity selects relatively bright galaxies, therefore the sample 
is to be complete and more uniform in a larger range of redshifts as compared 
to flux-limited samples.

We do not apply completeness corrections for the LF faint-end because the
deviations from the Schechter function were found only for the LF bright-end
which is not affected by the sample incompleteness. 

Can the difference be a result of the LF evolution with the redshift because
of the relatively large range of LCG redshifts of 0.02 -- 0.6?
The problem whether LF evolves with a redshift is still not resolved. Some 
authors found such dependence. However, for instance, 
\citet{I15} found that relations between global parameters for compact 
star-forming galaxies in much larger redshift range 0 -- 3 are universal. 
No dependence on the redshift and star-formation rate was found for 
mass-metallicity or luminosity-metallicity relation. This universality may 
imply the absence of the LF variations with the redshift for compact 
star-forming galaxies.

It is clear that the presence in galaxies of young stellar populations with 
luminous massive stars would result in an excess of galaxy LF as compared to 
that for quiescent galaxies with older stellar populations. This effect is 
stronger in the wavelength ranges, which trace star formation.
Furthermore, the differences between these LFs are not only quantitative but 
also qualitative in nature.  
The LF of the Schechter type, based on the Poisson distribution, is expected 
if the probability of the star formation in some small region of a galaxy does 
not depend on the star formation in surrounding regions. The deviation of the 
LF from the Schechter function indicates that the star-forming processes in the 
nearby regions correlate. Consequently, an excess of bright galaxies in the LF 
can be due to the propagating star formation stimulated by the burst.

An excess of LCGs on a bright end of LF is detected for the luminosities 
$L(\rmn{H}\alpha)>5\times 10^{42}$ erg s$^{-1}$. This is a 
characteristic luminosity of a strong starburst for LCG with the respective 
SFR $\ge 40 M_{\odot}$ yr$^{-1}$. The respective characteristic 
mass of the young stellar population of $m>2 \times 10^8 M_{\odot}$ is estimated
from the ratio $SFR/m \approx 1.95 \times 10^{-7}$, which was obtained for LCGs
by \citet{PII}.
 
\section{Conclusions}\label{s:concl}

We study the goodness of different functions for approximation of the H$\alpha$ and UV
luminosity functions (LFs) of luminous compact galaxies (LCGs). Some of these functions
take into account rapid luminosity evolution due to the presence of short-lived massive 
stars. We ruled out the Schechter function and the related
distribution (\ref{e15}), which badly fit the high-luminosity end. 
The Saunders function, the log-normal distribution
and functions (\ref{enn}), (\ref{enni}), derived in this paper, reproduce LFs much better. 
The choice of a preferable function depends on the wavelength.

The H$\alpha$ LF of the entire LCG sample is best fitted by either (\ref{enn},\ref{e11}) 
or (\ref{enni}) distribution
with similar corresponding probabilities (37\% and 36\%) of deviations being caused by 
random errors. The Saunders function with the 31\% probability is prefered compared 
to the log-normal distribution with the 27\% probability. 
The probability for the subsample of LCGs with a single knot of star formation is lower
compared to the entire sample. This implies that LCGs with several knots of 
star formation and those with a single knot have similar LFs with similar data scatter.

The function (\ref{enni}) is the best for fitting the FUV LF with corresponding 
probability of 94\%. However, the very low-luminosity end of the LF is better fitted by the 
log-normal function (Fig. \ref{FUV}). This fact does not affect the
$\chi^2$ value, because the $\chi^2$ test ignores a luminosity distribution inside the bins.

The NUV LF can be best fitted by the log-normal distribution with the lower confidence 
probability of 50\%, which is limited because of higher data scatter (Fig. \ref{NUV}). 

We demonstrate that the quality of fitting for two out of three LFs is considerably
increased by taking into account the rapid luminosity evolution. 
The improvement of the probability is 10\% for the H$\alpha$ LF and 33\% for the FUV LF.
The effect of rapid evolution is less pronounced for the NUV LF.
Using samples with larger ranges of luminosities would further improve the identification
of the best LF functional form.

The effect of the rapid luminosity evolution is more pronounced for 
the H$\alpha$ line 
than for the FUV and for the FUV than for the NUV as evidenced by the $q$ 
values. The optimal value provided by MLM and $\chi^2$-test for the NUV 
differs from ones for H$\alpha$ and FUV 
probably due to this fact. However, the minimum of $\chi^2$ the maximum of $U$ are
shallow, therefore errors of $q$ are large.

We assume that LFs for stellar populations with different
ages are different. We adopt that the LF for the emission in the optical 
continuum, produced by older stellar population, is described by the Schechter 
function. On the other hand, LFs of the emission in the UV and IR continua and 
the H$\alpha$ line, produced by young stellar populations, is described
by the log-normal distribution.
It is clear that the bright end of the LF of the young stellar population must 
exceed the LF of the older stellar population due to the contribution of
short-lived massive stars. 

Thus, starbursts producing a young stellar population with masses of 
at least $2 \times 10^8 M_{\odot}$ are 
more frequent than the Schechter function predicts. Such high masses
may be explained by the propagating star formation,
which is resulted in the enhanced luminosity
and, consequently, in an excess of LCGs at the bright end of LF in comparison 
with the case without an induced star formation.

There is no principal difference between the LFs of the whole sample and of 
the subsample of galaxies with a single star-forming knot.

\section*{Acknowledgments}
We are grateful to an anonymous referee for his/her
valuable comments. 
Funding for the Sloan Digital Sky Survey (SDSS) and SDSS-II has
been provided by the Alfred P. Sloan Foundation, the Participating
Institutions, the National Science Foundation, the U.S. Department
of Energy, the National Aeronautics and Space Administration, the
Japanese Monbukagakusho, and the Max Planck Society, and the
Higher Education Funding Council for England.

\label{lastpage}

\end{document}